\documentclass[aip,reprint]{revtex4-1}

\pdfoutput=1
\usepackage[charter]{mathdesign}
\usepackage{amsmath}
\usepackage{siunitx}
\usepackage[pdftex]{graphicx}
\usepackage{microtype}
\usepackage{bm}
\usepackage{siunitx}
\usepackage[version=3]{mhchem}

\usepackage[svgnames]{xcolor}
\usepackage[
	colorlinks=True,linkcolor=DarkRed,citecolor=ForestGreen,urlcolor=MediumBlue,
	pdfstartview=FitH,bookmarks=False,pdfpagemode=UseNone
]{hyperref}

\newcommand{\Tg}{$T_\textrm{g}$}
\newcommand{\mob}[1]{\SI{#1}{\square \centi \meter \per \volt \per \second}}
\newcommand{\cdens}[1]{\SI{#1}{\per \square \centi \meter}}
\newcommand{\Rxy}{$R_{xy}(B)$}
\newcommand{\Rs}{$R_\textrm{s}$}
\newcommand{\dxy}{$d_{xy}$}
\newcommand{\dxz}{$d_{xz}$}
\newcommand{\dyz}{$d_{yz}$}
\newcommand{\dxzyz}{$d_{xz}/d_{yz}$}

\begin{document}

\title{\boldmath Growth-induced electron mobility enhancement at the \ce{LaAlO3}/\ce{SrTiO3} interface}

\author{A. F{\^e}te}
\affiliation{D{\'e}partement de Physique de la Mati{\`e}re Condens{\'e}e, Universit{\'e} de Gen{\`e}ve, 24 Quai Ernest-Ansermet, 1211 Gen{\`e}ve 4, Suisse}
\author{C. Cancellieri}
\affiliation{D{\'e}partement de Physique de la Mati{\`e}re Condens{\'e}e, Universit{\'e} de Gen{\`e}ve, 24 Quai Ernest-Ansermet, 1211 Gen{\`e}ve 4, Suisse}
\affiliation{Now at Empa, Swiss Federal Laboratories for Materials Testing and Research, \"Uberlandstrasse 129, 8600 D\"ubendorf, Switzerland}
\author{D. Li}
\affiliation{D{\'e}partement de Physique de la Mati{\`e}re Condens{\'e}e, Universit{\'e} de Gen{\`e}ve, 24 Quai Ernest-Ansermet, 1211 Gen{\`e}ve 4, Suisse}
\author{D. Stornaiuolo}
\affiliation{D{\'e}partement de Physique de la Mati{\`e}re Condens{\'e}e, Universit{\'e} de Gen{\`e}ve, 24 Quai Ernest-Ansermet, 1211 Gen{\`e}ve 4, Suisse}
\affiliation{Now at Dipartimento di Fisica, Universit{\`a} degli Studi di Napoli Federico II, 80126 Napoli, Italy}
\author{A.  D. Caviglia}
\affiliation{D{\'e}partement de Physique de la Mati{\`e}re Condens{\'e}e, Universit{\'e} de Gen{\`e}ve, 24 Quai Ernest-Ansermet, 1211 Gen{\`e}ve 4, Suisse}
\affiliation{Now at Kavli Institute of Nanoscience, Delft University of Technology, P.O. Box 5046, 2600 GA Delft, The Netherlands}
\author{S. Gariglio}
\affiliation{D{\'e}partement de Physique de la Mati{\`e}re Condens{\'e}e, Universit{\'e} de Gen{\`e}ve, 24 Quai Ernest-Ansermet, 1211 Gen{\`e}ve 4, Suisse}
\author{J.-M. Triscone}
\affiliation{D{\'e}partement de Physique de la Mati{\`e}re Condens{\'e}e, Universit{\'e} de Gen{\`e}ve, 24 Quai Ernest-Ansermet, 1211 Gen{\`e}ve 4, Suisse}

\begin{abstract}
We have studied the electronic properties of the 2D electron liquid present at the \ce{LaAlO3}/\ce{SrTiO3} interface in series of samples prepared at different growth temperatures. We observe that interfaces fabricated at \SI{650}{\celsius} exhibit the highest low temperature mobility ($\approx \mob{10000}$) and the lowest sheet carrier density ($\approx \cdens{5e12}$). These samples show metallic behavior and Shubnikov-de Haas oscillations in their magnetoresistance. Samples grown at higher temperatures ($800-\SI{900}{\celsius}$) display carrier densities in the range of $\approx 2-\cdens{5e13}$ and mobilities of $\approx \mob{1000}$ at \SI{4}{\kelvin}.  Reducing their carrier density by field effect to $\cdens{8e12}$ lowers their mobilites to $\approx \mob{50}$ bringing the conductance to the weak-localization regime.
\end{abstract}
\maketitle

Transition metal oxide heterostructures are very promising for nano-electronics since they allow the realization and tuning of new electronic phases \cite{Takagi2010,Mannhart2010,Zubko2011}. In this respect, the interface formed by the two insulating perovskites \ce{LaAlO3} (LAO) and \ce{SrTiO3} (STO) is attracting a lot of attention due to the presence of a two-dimensional electron liquid (2DEL) \cite{Ohtomo2004}, whose density can be tuned by electric field effect revealing a rich phase diagram \cite{Thiel2006,Caviglia2008}. The origin of conduction is thought to be linked to the polar discontinuity between the two materials \cite{Nakagawa2006,Bristowe2011,Cancellieri2011,Stengel2011}.

A key ingredient for electronic applications is the charge mobility $\mu$. LAO/STO interfaces typically display mobilities of \mob{1000} at low temperature \cite{Reyren2007,Huijben2009}, a value relatively low if electrons are provided by charge transfer. Hence, different studies have attempted to pinpoint the origin of the scattering mechanisms that limit the electron mobility at the interface. Thiel \textit{et al.} observed a dramatic reduction of the conductance of the 2DEL when the density of dislocations is increased, and suggested that the region of influence of these line defects extends well beyond their physical size \cite{Thiel2009}. In a related work Fix and coworkers considered the effect of vicinal substrates (i.e. step edge density) on mobility \cite{Fix2011}. In agreement with previous results \cite{Bell2009a}, the proximity of the charge distribution to the interface was pointed out to be an important factor in determining the scattering rate. Similar considerations were also raised for heterostructures of \ce{LaTiO3}/\ce{SrTiO3} where the electric field effect was used to modify the charge profile \cite{Kim2010}. The role of the LAO thickness on the electron mobility at the LAO/STO interface was also studied, revealing a strongly reduced mobility in heterostructures made of  thick LAO films \cite{Bell2009}. More recently, the presence of tetragonal domains in STO (at low temperature) was demonstrated to alter locally the conductance \cite{Kalisky2013,Honig2013}. Finally, the control of the LAO surface state was shown to be an efficient lever to enhance the interfacial electron mobility \cite{Xie2013,Huijben2013,Bristowe2011}.

In this letter, we perform a systematic study of the effect of the growth temperature (\Tg) on the transport properties of the LAO/STO interface. 
In samples grown at $800-\SI{900}{\celsius}$, called ``standard'' samples, carrier densities in the $2-\cdens{6e13}$ range are observed while mobility at low temperature reaches $\approx \mob{1000}$. Reducing the growth temperature to \Tg \ $\approx \SI{650}{\celsius}$ leads to a lower carrier density, of the order of a few \cdens{e12}, and a much higher mobility reaching $\mob{8000}$, allowing the observation of 2D-quantum oscillations \cite{Caviglia2010105}. Concomitantly, whereas the Hall resistance \Rxy \ is found to be temperature dependent and to display a non-linear behavior at low temperature for ``standard'' samples, \Rxy \ is $T$ independent and linear for interfaces grown at \SI{650}{\celsius}. 
We also show that, using the electric field effect to reduce the Hall carrier density of high \Tg \ samples (having initial high carrier densities) to values observed in low \Tg \ samples, the magnetoresistance differs from the one observed for high-mobility samples.

LAO thin films were grown by pulsed laser deposition at an oxygen pressure of \SI{8e-5}{\milli \bar} and post-annealed for one hour at \SI{530}{\celsius} in \SI{200}{\milli \bar} of oxygen \cite{Cancellieri2010}. The KrF excimer laser was set for a fluence of \SI{0.6}{\joule \per \square \centi \meter} with a repetition rate of \SI{1}{\hertz}. Growth temperatures \Tg \ ranged from \SI{650}{\celsius} to \SI{900}{\celsius}, as measured on the substrate by an optical pyrometer. The thickness of the LAO layer was kept within the range of 5 to 10 unit cells \cite{Cancellieri2011}.
Periodic oscillations in the reflection high energy electron diffraction (RHEED) intensity reveal a layer-by-layer growth for all \Tg. The RHEED diffraction pattern indicates that the 2-dimensional character is maintained even for the lowest \Tg. Atomic force microscopy confirms that the film surfaces reproduce the step-and-terrace morphology of the TiO$_2$-terminated (001) STO substrates \footnote{We use STO substrates from CrysTec GmbH that are already HF treated and high-temperature annealed to provide \ce{TiO2} termination. Before growth, the sample holder with the substrate is annealed in vacuum for 30 min at a temperature of \SI{290}{\celsius}.}.
X-ray $\theta-2\theta$ scans for the 001 reflections show diffraction peaks with finite size fringes in agreement with the layer thickness and an intensity independent of \Tg. The above structural characterization revealed no effect of the substrate temperature on the growth mode or the crystalline quality of thin LAO layers.

For DC transport measurements, \SI{500}{\micro \meter} wide Hall bridges were patterned using standard photolithography techniques and an amorphous STO hard mask \cite{Stornaiuolo2012}.

%%%%%%%%%%%%%%%%%%%%%%%%%%%%%%%%%%%%%%%%%%%%%%%%%%%%%%
%%%%%%%%%%%%%%%%%%%%%%%%%%%%%%%%%%%%%%%%%%%%%%%%%%%%%%
%%%%%%%%%%%%%%%%%%%%%%%%%%%%%%%%%%%%%%%%%%%%%%%%%%%%%%
\begin{figure}
\centering\includegraphics[width=1\columnwidth]{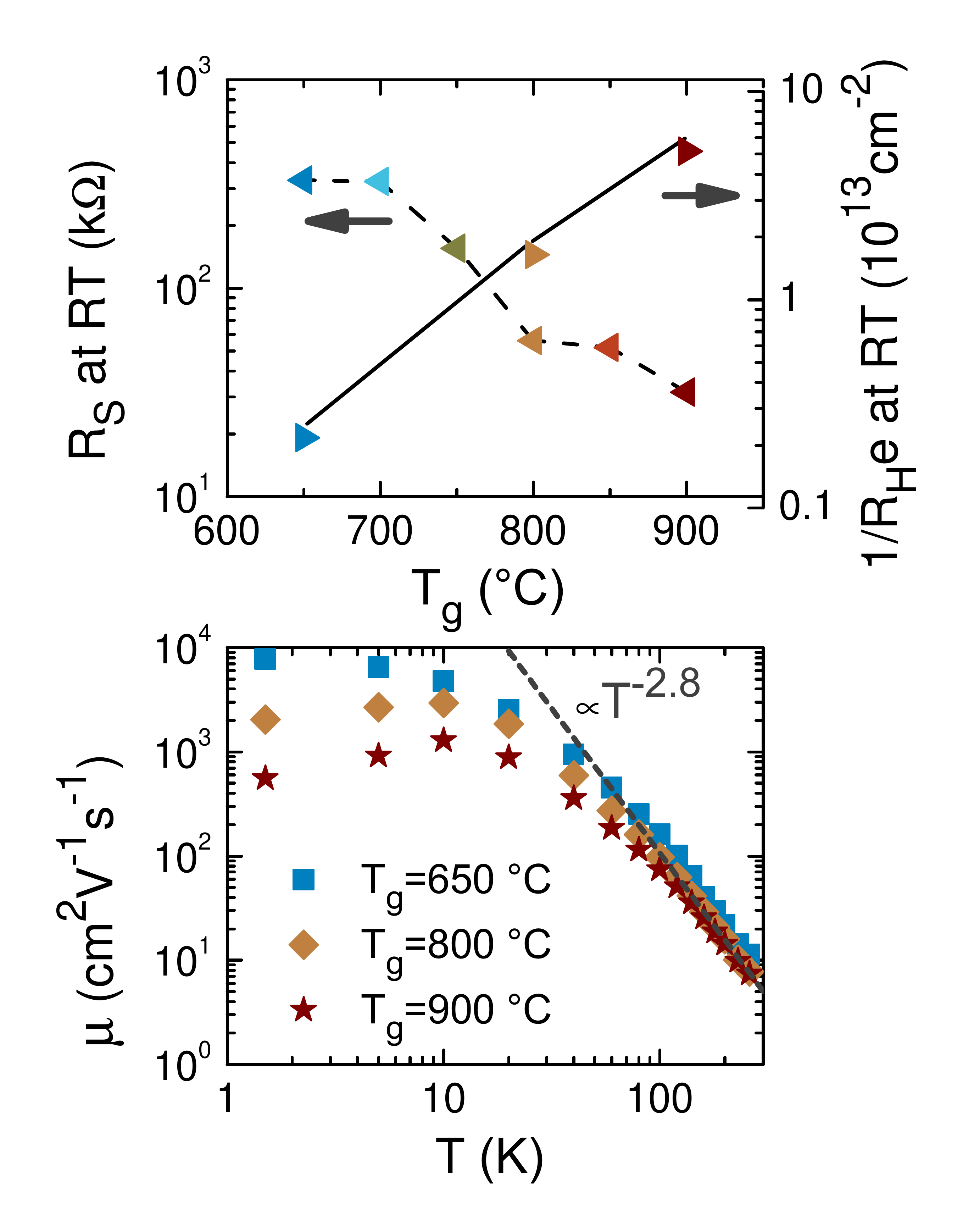}
\caption{\label{Fig1}
Top panel: Evolution of the sheet resistance (left scale) and inverse Hall constant (right scale) measured at RT as a function of the growth temperature \Tg. RT stands for \SI{296}{\kelvin} and \SI{260}{\kelvin} for the resistance and Hall measurements respectively. The lines are guides to the eye. 
Middle panel: Hall mobility as a function of temperature for samples grown at \SI{900}{\celsius} (brown stars), \SI{800}{\celsius} (gold diamonds) and \SI{650}{\celsius} (blue squares). The dashed line shows a fit assuming $\mu \propto T^\alpha$ with $\alpha$ being $-2.8$. This temperature dependence is also observed in bulk doped-STO.
}
\end{figure}
%%%%%%%%%%%%%%%%%%%%%%%%%%%%%%%%%%%%%%%%%%%%%%%%%%%%%%
%%%%%%%%%%%%%%%%%%%%%%%%%%%%%%%%%%%%%%%%%%%%%%%%%%%%%%
%%%%%%%%%%%%%%%%%%%%%%%%%%%%%%%%%%%%%%%%%%%%%%%%%%%%%%

In the top panel of Fig.~\ref{Fig1}, the room temperature (RT) sheet resistance \Rs (\SI{296}{\kelvin}) is plotted as a function of the growth temperature. We note that upon decreasing \Tg \ down to \SI{650}{\celsius} the RT sheet resistance \Rs \ increases by roughly an order of magnitude. This evolution is mainly related to a reduction of the carrier density, the increase in the Hall mobility being less than a factor of two \footnote{The estimation of the carrier density is obtained from the Hall constant at \SI{296}{\kelvin} where the Hall resistance is linear.}. The lower panel of Fig.~\ref{Fig1} shows a large increase in the electron mobility at low temperatures : as \Tg \ is reduced from 900 to \SI{650}{\celsius}, $\mu$ goes from 600 to \mob{8000} at \SI{1.5}{\kelvin}.

%%%%%%%%%%%%%%%%%%%%%%%%%%%%%%%%%%%%%%%%%%%%%%%%%%%%%
%%%%%%%%%%%%%%%%%%%%%%%%%%%%%%%%%%%%%%%%%%%%%%%%%%%%%%
%%%%%%%%%%%%%%%%%%%%%%%%%%%%%%%%%%%%%%%%%%%%%%%%%%%%%%
\begin{figure*}
\centering\includegraphics[width=1.55\columnwidth]{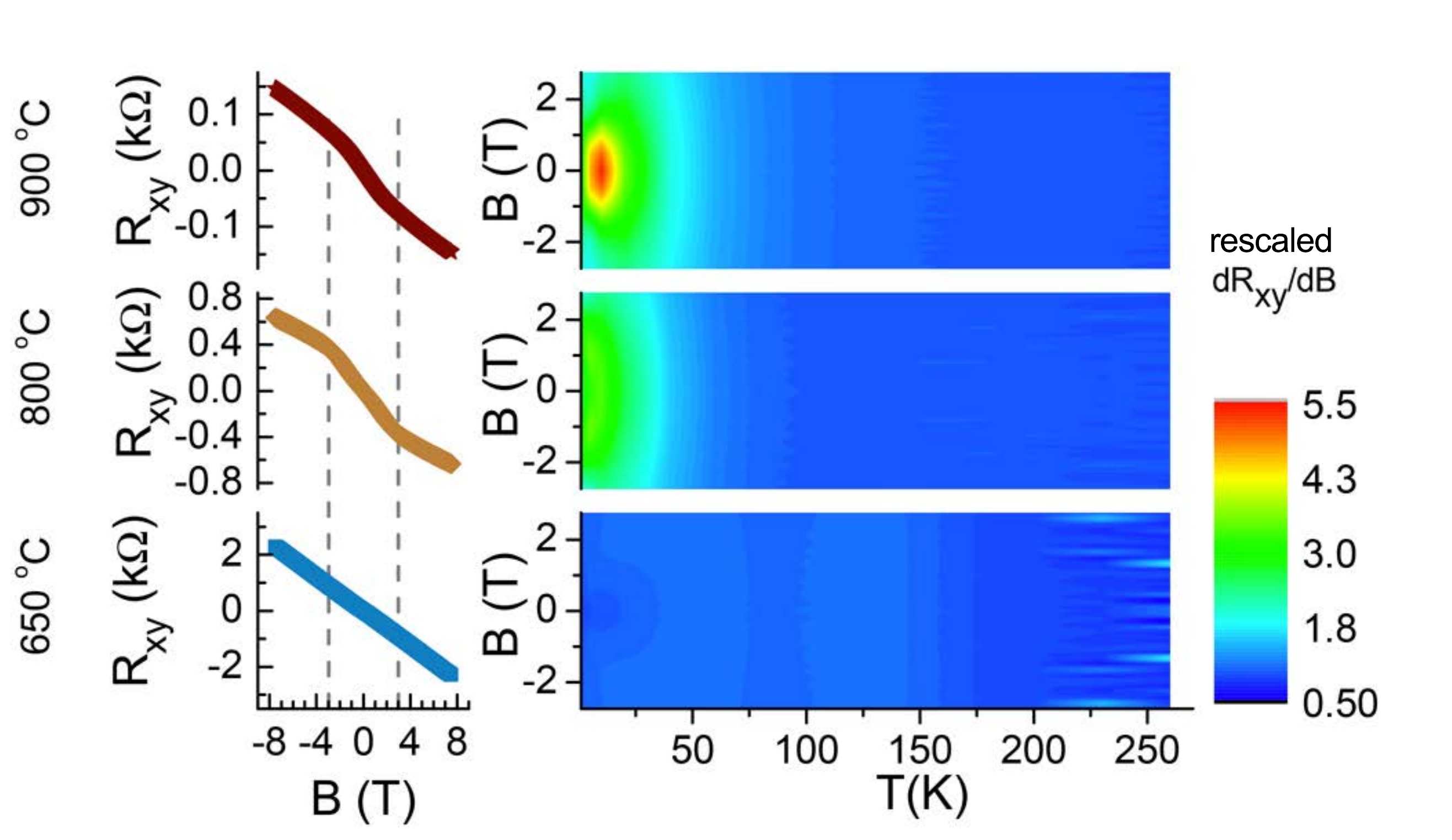}
\caption{\label{Fig2}
(left) \Rxy \ vs. magnetic field $B$ measured at \SI{1.5}{\kelvin} for three growth temperatures. (right) First derivative of \Rxy \  as a function of magnetic field and temperature. For each growth temperature, the value of the derivative has been rescaled by its value at \SI{160}{\kelvin} $(\partial R_{xy}/\partial B)/(\partial R_{xy}/\partial B)|_{T=\SI{160}{\kelvin}}$ to highlight the presence of non-linearities.}
\end{figure*}
%%%%%%%%%%%%%%%%%%%%%%%%%%%%%%%%%%%%%%%%%%%%%%%%%%%%%%
%%%%%%%%%%%%%%%%%%%%%%%%%%%%%%%%%%%%%%%%%%%%%%%%%%%%%%
%%%%%%%%%%%%%%%%%%%%%%%%%%%%%%%%%%%%%%%%%%%%%%%%%%%%%%

In Fig.~\ref{Fig2}, we plot (left) the Hall resistance \Rxy \  measured at \SI{1.5}{\kelvin} as a function of the magnetic field and (right) its field derivative as a function of the magnetic field $B$ and temperature $T$ for heterostructures grown at three different temperatures.

\begin{figure}
\centering
\includegraphics[width=0.8\columnwidth]{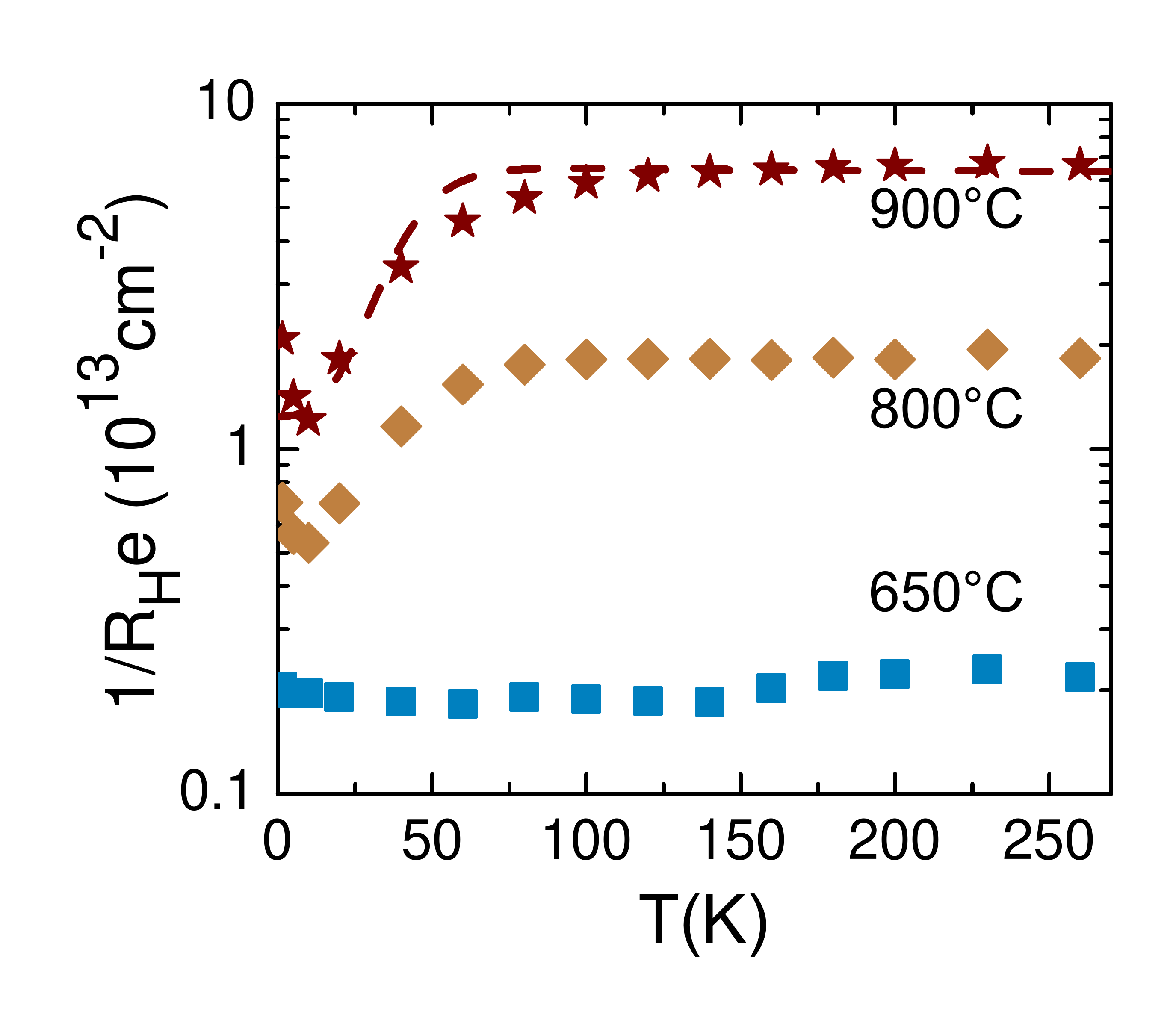}
\caption{\label{Fig3}
Temperature dependence of the inverse of the Hall constant measured at $B \rightarrow 0$ (defined as $\partial R_{xy}/\partial B |_{B\rightarrow 0}$) for interfaces grown at different temperatures (symbols) and calculated using the model described in the text (dashed line).}
\end{figure}

Samples grown at \Tg=\SI{650}{\celsius} exhibit a linear \Rxy \  essentially temperature independent. For growth temperatures of 800 and \SI{900}{\celsius}, the Hall resistance \Rxy \  displays a more complex behavior. In the 100-\SI{300}{\kelvin} temperature range, \Rxy \  is linear in magnetic field and its derivative ($\partial R_{xy}/\partial B$) is temperature independent \footnote{We note, though, that depending on the photo-doping level \cite{Tebano2012,Guduru2013}, some temperature dependence of the Hall effect above \SI{100}{\kelvin} can sometimes be observed.}. Below $\approx \SI{100}{\kelvin}$, the Hall response becomes a non-linear function of the magnetic field and its low field derivative increases at low temperature. The temperature dependence of the inverse Hall constant estimated at $B \rightarrow 0$ is shown in Fig.~\ref{Fig3}.

Analyzing the low temperature magnetotransport for samples grown at \SI{900}{\celsius} with a two channel conduction model reveals contributions from a band with a low carrier density ($1-\cdens{5e12}$) and a high mobility (in the range of $1000-\mob{2000}$) and from a band with a higher carrier density ($\approx 5-\cdens{6e13}$) and low mobility (in the range of $100-\mob{200}$), in agreement with previous reports \cite{BenShalom2010105,Lerer2011}.
The dependence on magnetic field of the low field $R_\textrm{H}$ for the different temperatures can be captured by the same two-band model (see Fig.~\ref{Fig3}) taking into account the temperature dependence of the scattering time ($1/\tau=1/\tau_\textrm{e}+1/\tau_\textrm{i}$, $\tau_\textrm{e}$ and $\tau_\textrm{i}$ being the elastic and the inelastic scattering times respectively and $\tau_\textrm{i}\propto T^{-2.8}$, see Fig.~\ref{Fig1}) and keeping the density of each band constant \footnote{The elastic scattering time for each band is calculated from the respective mobility obtained from the two band fit in magnetic field while the inelastic scattering time is considered to be independent off the carrier type.}. Thus the temperature and field dependencies of the Hall resistance can be quantitatively explained by the presence of carriers with different mobilities in the system. This analysis also suggests that the correct estimate of the total mobile carrier density can be obtained from the slope of the Hall effect at high fields in the low $T$ range or above \SI{100}{\kelvin} \cite{Takahashi2006}.

%%%%%%%%%%%%%%%%%%%%%%%%%%%%%%%%%%%%%%%%%%%%%%%%%%%%%
%%%%%%%%%%%%%%%%%%%%%%%%%%%%%%%%%%%%%%%%%%%%%%%%%%%%%%
%%%%%%%%%%%%%%%%%%%%%%%%%%%%%%%%%%%%%%%%%%%%%%%%%%%%%%
%%%%%%%%%%%%%%%%%%%%%%%%%%%%%%%%%%%%%%%%%%%%%%%%%%%%%%
%%%%%%%%%%%%%%%%%%%%%%%%%%%%%%%%%%%%%%%%%%%%%%%%%%%%%%
%%%%%%%%%%%%%%%%%%%%%%%%%%%%%%%%%%%%%%%%%%%%%%%%%%%%%%

The presence of charge carriers with different masses (and thus different mobilities) at the LAO/STO interface has been anticipated by \textit{ab initio} calculations \cite{Pentcheva2006,Popovic2008,Son2009,Delugas2011,Stengel2011}. The predicted electronic structure of the confined electron liquid is characterized by a lifting of the $t_{2g}$ degeneracy of the 3$d$ orbitals with the lowest \dxy \  sub-band(s)  having  a lower energy than the lowest \dxzyz \  sub-bands. Experimental evidence supporting this model came from linear dichroism measurements \cite{Salluzzo2009} providing an estimate of their splitting of $\approx\SI{50}{\milli \electronvolt}$. According to this picture, for large densities, the Fermi level should cross sub-bands with \dxz \ and \dyz \ ($m^{*}=2.2 m_\textrm{e}$, $m_\textrm{e}$ being the electron mass) character and then multi-channel conduction is expected. For low carrier densities only (a) \dxy \  sub-band(s) with an effective mass $m^{*}=0.7 m_\textrm{e}$ should contribute to the transport. However, more recently resonant photoemission spectroscopy\cite{Cancellieri2014} supported by \textit{ab initio} calculations have revealed that the electronic band structure is modified along with the carrier density, probably related to a change in the confinement potential; \dxzyz \ sub-bands are still observed at low carrier density interfaces.

%%%%%%%%%%%%%%%%%%%%%%%%%%%%%%%%%%%%%%%%%%%%%%%%%%%%%
%%%%%%%%%%%%%%%%%%%%%%%%%%%%%%%%%%%%%%%%%%%%%%%%%%%%%%
%%%%%%%%%%%%%%%%%%%%%%%%%%%%%%%%%%%%%%%%%%%%%%%%%%%%%%
\begin{figure}
\centering\includegraphics[width=0.8\columnwidth]{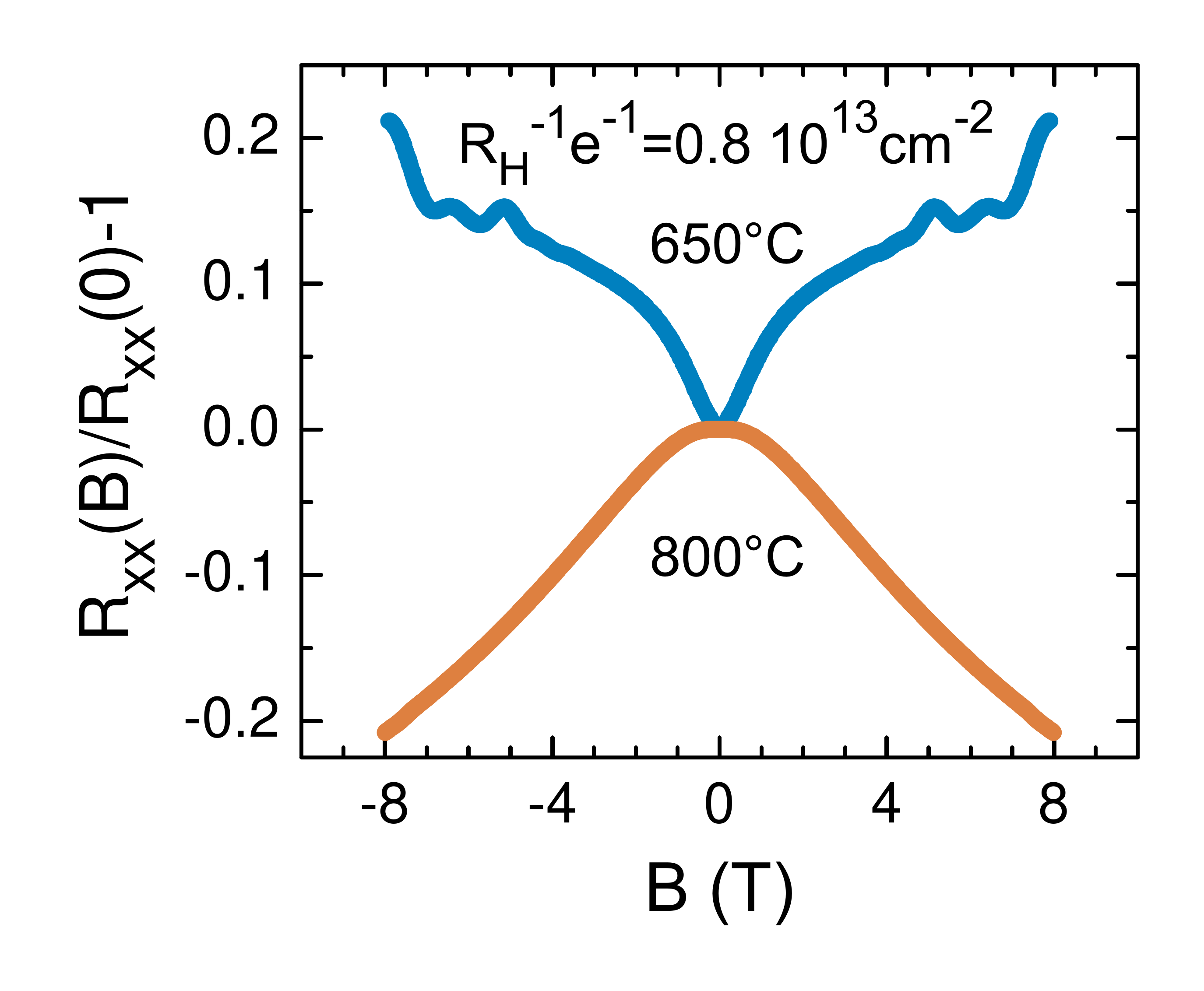}
\caption{\label{Fig4}
Comparison between the magnetotransport of two interfaces exhibiting the same sheet carrier densities but grown at different temperatures. For the interface grown at \SI{800}{\celsius}, the electric field effect was used to reduce the carrier density. A low value for the mobility (\mob{34}) is measured and analyses of magnetotransport reveal a weak localization regime. For the one grown at \SI{650}{\celsius} Shubnikov-de Haas oscillations can be observed revealing high-mobility.
}
\end{figure}
%%%%%%%%%%%%%%%%%%%%%%%%%%%%%%%%%%%%%%%%%%%%%%%%%%%%%%
%%%%%%%%%%%%%%%%%%%%%%%%%%%%%%%%%%%%%%%%%%%%%%%%%%%%%%
%%%%%%%%%%%%%%%%%%%%%%%%%%%%%%%%%%%%%%%%%%%%%%%%%%%%%%

Fig.~\ref{Fig4} shows the change in magnetoresistance at \SI{1.5}{\kelvin} for a sample grown at \SI{650}{\celsius} ($\mu=\mob{5050}$): it amounts to 20\% for a magnetic field of \SI{7}{\tesla}. Above \SI{4}{\tesla}, we observe the appearance of Shubnikov-de Haas oscillations. This large magnetoresistance suggests that also for these low carrier densities multiple channels contribute to the transport. Further evidence of the population of \dxzyz \ orbitals in  LAO/STO interfaces grown at \SI{650}{\celsius} were provided recently by detailed analyses of the Shubnikov-de Haas oscillations \cite{Fete2014}.

In order to shed further light on the effect of the carrier density on the sample mobility, we performed a series of field effect experiments in samples grown at \SI{800}{\celsius}. In agreement with Bell and coworkers \cite{Bell2009a}, we find that the dominant effect of the electric field in the back gate geometry is the modulation of the mobility. In Fig.~\ref{Fig4} we show also the low temperature magnetoresistance of a sample grown at \SI{800}{\celsius}. Using field effect, its carrier density was tuned to match that of the high mobility sample \footnote{In this carrier density range, the Hall effect is linear in magnetic field for both type of interfaces.} ($\textrm{R}_\textrm{H}^{-1} e^{-1}= \cdens{8e12}$). As a result, the mobility was reduced to \mob{34}. Fig.~\ref{Fig4} shows that whereas the high mobility sample displays a positive magnetoresistance, the change in resistance of the \SI{800}{\celsius} sample is negative and on the scale of $e^2/\pi h$. This behavior has been shown to be the signature of weak-localization \cite{Caviglia2010}.
This comparison reveals strikingly different magnetotransport behavior in samples with similar carrier densities and confirm the low dimensional nature of this conducting system where transport regimes are determined by the characteristic scattering lengths (elastic, phase, spin-orbit) more than their carrier density.

These results emphasize the open question on the origin of the carrier mobility at the LAO/STO interface : what is determining the high $\mu$ observed in low \Tg \ samples compared with the low $\mu$ measured in high \Tg \ samples for the same doping ? The shape of the confinement potential is responsible for the proximity of the conducting layer to the interface. Hence, any disorder, as for instance interface intermixing, would affect differently the electron mobility depending on the charge profile \cite{Bell2009a}. In this scenario, the electric field effect results suggest that the charge profile at the same density can be different. On the other hand, if interface perfection is modified by the growth temperature \cite{Ohnishi2004}, then the same charge profile would generate a different electronic mobility.

Both situations can be at play in our samples; further detailed structural analyses, like surface x-ray diffraction, could help to address this important point.

In conclusion, transport properties of \ce{LaAlO3}/\ce{SrTiO3} interfaces were shown to be very sensitive to the sample growth temperature. For a substrate temperature of \SI{650}{\celsius}, the carrier mobility reaches \mob{8000} and Shubnikov-de Haas oscillations appear in the magnetotransport.  These samples also show a low carrier density of $\approx \cdens{2e12}$. Field effect experiments reveal that the carrier density is not the parameter determining the mobility. Further optimization of the growth conditions remains key to the observation of clean superconductivity and other quantum effects in this complex oxide 2D electron system.

The authors are grateful to M.~Lopes and
S.~C.~M{\"u}ller for their technical assistance. This work was supported by the
Swiss National Science Foundation through the NCCR MaNEP and Division II, by
the Institut Universitaire de France (MG) and has received funding from the European Research Council under the European Union's Seventh Framework Programme (FP7/2007-2013) / ERC Grant Agreement n$^\circ$ 319286 (Q-MAC).

\bibliographystyle{apsrev4-1}

\begin{thebibliography}{41}%
\makeatletter
\providecommand \@ifxundefined [1]{%
 \@ifx{#1\undefined}
}%
\providecommand \@ifnum [1]{%
 \ifnum #1\expandafter \@firstoftwo
 \else \expandafter \@secondoftwo
 \fi
}%
\providecommand \@ifx [1]{%
 \ifx #1\expandafter \@firstoftwo
 \else \expandafter \@secondoftwo
 \fi
}%
\providecommand \natexlab [1]{#1}%
\providecommand \enquote  [1]{``#1''}%
\providecommand \bibnamefont  [1]{#1}%
\providecommand \bibfnamefont [1]{#1}%
\providecommand \citenamefont [1]{#1}%
\providecommand \href@noop [0]{\@secondoftwo}%
\providecommand \href [0]{\begingroup \@sanitize@url \@href}%
\providecommand \@href[1]{\@@startlink{#1}\@@href}%
\providecommand \@@href[1]{\endgroup#1\@@endlink}%
\providecommand \@sanitize@url [0]{\catcode `\\12\catcode `\$12\catcode
  `\&12\catcode `\#12\catcode `\^12\catcode `\_12\catcode `\%12\relax}%
\providecommand \@@startlink[1]{}%
\providecommand \@@endlink[0]{}%
\providecommand \url  [0]{\begingroup\@sanitize@url \@url }%
\providecommand \@url [1]{\endgroup\@href {#1}{\urlprefix }}%
\providecommand \urlprefix  [0]{URL }%
\providecommand \Eprint [0]{\href }%
\providecommand \doibase [0]{http://dx.doi.org/}%
\providecommand \selectlanguage [0]{\@gobble}%
\providecommand \bibinfo  [0]{\@secondoftwo}%
\providecommand \bibfield  [0]{\@secondoftwo}%
\providecommand \translation [1]{[#1]}%
\providecommand \BibitemOpen [0]{}%
\providecommand \bibitemStop [0]{}%
\providecommand \bibitemNoStop [0]{.\EOS\space}%
\providecommand \EOS [0]{\spacefactor3000\relax}%
\providecommand \BibitemShut  [1]{\csname bibitem#1\endcsname}%
\let\auto@bib@innerbib\@empty
%</preamble>
\bibitem [{\citenamefont {Takagi}\ and\ \citenamefont
  {Hwang}(2010)}]{Takagi2010}%
  \BibitemOpen
  \bibfield  {author} {\bibinfo {author} {\bibfnamefont {H.}~\bibnamefont
  {Takagi}}\ and\ \bibinfo {author} {\bibfnamefont {H.~Y.}\ \bibnamefont
  {Hwang}},\ }\href {\doibase 10.1126/science.1182541} {\bibfield  {journal}
  {\bibinfo  {journal} {Science}\ }\textbf {\bibinfo {volume} {327}},\ \bibinfo
  {pages} {1601} (\bibinfo {year} {2010})}\BibitemShut {NoStop}%
\bibitem [{\citenamefont {Mannhart}\ and\ \citenamefont
  {Schlom}(2010)}]{Mannhart2010}%
  \BibitemOpen
  \bibfield  {author} {\bibinfo {author} {\bibfnamefont {J.}~\bibnamefont
  {Mannhart}}\ and\ \bibinfo {author} {\bibfnamefont {D.~G.}\ \bibnamefont
  {Schlom}},\ }\href {\doibase 10.1126/science.1181862} {\bibfield  {journal}
  {\bibinfo  {journal} {Science}\ }\textbf {\bibinfo {volume} {327}},\ \bibinfo
  {pages} {1607} (\bibinfo {year} {2010})}\BibitemShut {NoStop}%
\bibitem [{\citenamefont {Zubko}\ \emph {et~al.}(2011)\citenamefont {Zubko},
  \citenamefont {Gariglio}, \citenamefont {Gabay}, \citenamefont {Ghosez},\
  and\ \citenamefont {Triscone}}]{Zubko2011}%
  \BibitemOpen
  \bibfield  {author} {\bibinfo {author} {\bibfnamefont {P.}~\bibnamefont
  {Zubko}}, \bibinfo {author} {\bibfnamefont {S.}~\bibnamefont {Gariglio}},
  \bibinfo {author} {\bibfnamefont {M.}~\bibnamefont {Gabay}}, \bibinfo
  {author} {\bibfnamefont {P.}~\bibnamefont {Ghosez}}, \ and\ \bibinfo {author}
  {\bibfnamefont {J.-M.}\ \bibnamefont {Triscone}},\ }\href {\doibase
  10.1146/annurev-conmatphys-062910-140445} {\bibfield  {journal} {\bibinfo
  {journal} {Annu. Rev. Condens. Matter Phys.}\ }\textbf {\bibinfo {volume}
  {2}},\ \bibinfo {pages} {141} (\bibinfo {year} {2011})}\BibitemShut {NoStop}%
\bibitem [{\citenamefont {Ohtomo}\ and\ \citenamefont
  {Hwang}(2004)}]{Ohtomo2004}%
  \BibitemOpen
  \bibfield  {author} {\bibinfo {author} {\bibfnamefont {A.}~\bibnamefont
  {Ohtomo}}\ and\ \bibinfo {author} {\bibfnamefont {H.~Y.}\ \bibnamefont
  {Hwang}},\ }\href {\doibase 10.1038/nature02308} {\bibfield  {journal}
  {\bibinfo  {journal} {Nature}\ }\textbf {\bibinfo {volume} {427}},\ \bibinfo
  {pages} {423} (\bibinfo {year} {2004})}\BibitemShut {NoStop}%
\bibitem [{\citenamefont {Thiel}\ \emph {et~al.}(2006)\citenamefont {Thiel},
  \citenamefont {Hammerl}, \citenamefont {Schmehl}, \citenamefont {Schneider},\
  and\ \citenamefont {Mannhart}}]{Thiel2006}%
  \BibitemOpen
  \bibfield  {author} {\bibinfo {author} {\bibfnamefont {S.}~\bibnamefont
  {Thiel}}, \bibinfo {author} {\bibfnamefont {G.}~\bibnamefont {Hammerl}},
  \bibinfo {author} {\bibfnamefont {A.}~\bibnamefont {Schmehl}}, \bibinfo
  {author} {\bibfnamefont {C.~W.}\ \bibnamefont {Schneider}}, \ and\ \bibinfo
  {author} {\bibfnamefont {J.}~\bibnamefont {Mannhart}},\ }\href {\doibase
  10.1126/science.1131091} {\bibfield  {journal} {\bibinfo  {journal}
  {Science}\ }\textbf {\bibinfo {volume} {313}},\ \bibinfo {pages} {1942}
  (\bibinfo {year} {2006})}\BibitemShut {NoStop}%
\bibitem [{\citenamefont {Caviglia}\ \emph {et~al.}(2008)\citenamefont
  {Caviglia}, \citenamefont {Gariglio}, \citenamefont {Reyren}, \citenamefont
  {Jaccard}, \citenamefont {Schneider}, \citenamefont {Gabay}, \citenamefont
  {Thiel}, \citenamefont {Hammerl}, \citenamefont {Mannhart},\ and\
  \citenamefont {Triscone}}]{Caviglia2008}%
  \BibitemOpen
  \bibfield  {author} {\bibinfo {author} {\bibfnamefont {A.~D.}\ \bibnamefont
  {Caviglia}}, \bibinfo {author} {\bibfnamefont {S.}~\bibnamefont {Gariglio}},
  \bibinfo {author} {\bibfnamefont {N.}~\bibnamefont {Reyren}}, \bibinfo
  {author} {\bibfnamefont {D.}~\bibnamefont {Jaccard}}, \bibinfo {author}
  {\bibfnamefont {T.}~\bibnamefont {Schneider}}, \bibinfo {author}
  {\bibfnamefont {M.}~\bibnamefont {Gabay}}, \bibinfo {author} {\bibfnamefont
  {S.}~\bibnamefont {Thiel}}, \bibinfo {author} {\bibfnamefont
  {G.}~\bibnamefont {Hammerl}}, \bibinfo {author} {\bibfnamefont
  {J.}~\bibnamefont {Mannhart}}, \ and\ \bibinfo {author} {\bibfnamefont
  {J.-M.}\ \bibnamefont {Triscone}},\ }\href {\doibase 10.1038/nature07576}
  {\bibfield  {journal} {\bibinfo  {journal} {Nature}\ }\textbf {\bibinfo
  {volume} {456}},\ \bibinfo {pages} {624} (\bibinfo {year}
  {2008})}\BibitemShut {NoStop}%
\bibitem [{\citenamefont {Nakagawa}\ \emph {et~al.}(2006)\citenamefont
  {Nakagawa}, \citenamefont {Hwang},\ and\ \citenamefont
  {Muller}}]{Nakagawa2006}%
  \BibitemOpen
  \bibfield  {author} {\bibinfo {author} {\bibfnamefont {N.}~\bibnamefont
  {Nakagawa}}, \bibinfo {author} {\bibfnamefont {H.~Y.}\ \bibnamefont {Hwang}},
  \ and\ \bibinfo {author} {\bibfnamefont {D.~A.}\ \bibnamefont {Muller}},\
  }\href {\doibase 10.1038/nmat1569} {\bibfield  {journal} {\bibinfo  {journal}
  {Nat. Mater.}\ }\textbf {\bibinfo {volume} {5}},\ \bibinfo {pages} {204}
  (\bibinfo {year} {2006})}\BibitemShut {NoStop}%
\bibitem [{\citenamefont {Bristowe}\ \emph {et~al.}(2011)\citenamefont
  {Bristowe}, \citenamefont {Littlewood},\ and\ \citenamefont
  {Artacho}}]{Bristowe2011}%
  \BibitemOpen
  \bibfield  {author} {\bibinfo {author} {\bibfnamefont {N.~C.}\ \bibnamefont
  {Bristowe}}, \bibinfo {author} {\bibfnamefont {P.~B.}\ \bibnamefont
  {Littlewood}}, \ and\ \bibinfo {author} {\bibfnamefont {E.}~\bibnamefont
  {Artacho}},\ }\href {\doibase 10.1103/PhysRevB.83.205405} {\bibfield
  {journal} {\bibinfo  {journal} {Phys. Rev. B}\ }\textbf {\bibinfo {volume}
  {83}},\ \bibinfo {pages} {205405} (\bibinfo {year} {2011})}\BibitemShut
  {NoStop}%
\bibitem [{\citenamefont {Cancellieri}\ \emph {et~al.}(2011)\citenamefont
  {Cancellieri}, \citenamefont {Fontaine}, \citenamefont {Gariglio},
  \citenamefont {Reyren}, \citenamefont {Caviglia}, \citenamefont {Fete},
  \citenamefont {Leake}, \citenamefont {Pauli}, \citenamefont {Willmott},
  \citenamefont {Stengel}, \citenamefont {Ghosez},\ and\ \citenamefont
  {Triscone}}]{Cancellieri2011}%
  \BibitemOpen
  \bibfield  {author} {\bibinfo {author} {\bibfnamefont {C.}~\bibnamefont
  {Cancellieri}}, \bibinfo {author} {\bibfnamefont {D.}~\bibnamefont
  {Fontaine}}, \bibinfo {author} {\bibfnamefont {S.}~\bibnamefont {Gariglio}},
  \bibinfo {author} {\bibfnamefont {N.}~\bibnamefont {Reyren}}, \bibinfo
  {author} {\bibfnamefont {A.~D.}\ \bibnamefont {Caviglia}}, \bibinfo {author}
  {\bibfnamefont {A.}~\bibnamefont {Fete}}, \bibinfo {author} {\bibfnamefont
  {S.~J.}\ \bibnamefont {Leake}}, \bibinfo {author} {\bibfnamefont {S.~A.}\
  \bibnamefont {Pauli}}, \bibinfo {author} {\bibfnamefont {P.~R.}\ \bibnamefont
  {Willmott}}, \bibinfo {author} {\bibfnamefont {M.}~\bibnamefont {Stengel}},
  \bibinfo {author} {\bibfnamefont {P.}~\bibnamefont {Ghosez}}, \ and\ \bibinfo
  {author} {\bibfnamefont {J.-M.}\ \bibnamefont {Triscone}},\ }\href {\doibase
  10.1103/PhysRevLett.107.056102} {\bibfield  {journal} {\bibinfo  {journal}
  {Phys. Rev. Lett.}\ }\textbf {\bibinfo {volume} {107}},\ \bibinfo {pages}
  {56102} (\bibinfo {year} {2011})}\BibitemShut {NoStop}%
\bibitem [{\citenamefont {Stengel}(2011)}]{Stengel2011}%
  \BibitemOpen
  \bibfield  {author} {\bibinfo {author} {\bibfnamefont {M.}~\bibnamefont
  {Stengel}},\ }\href {\doibase 10.1103/PhysRevLett.106.136803} {\bibfield
  {journal} {\bibinfo  {journal} {Phys. Rev. Lett.}\ }\textbf {\bibinfo
  {volume} {106}},\ \bibinfo {pages} {136803} (\bibinfo {year}
  {2011})}\BibitemShut {NoStop}%
\bibitem [{\citenamefont {Reyren}\ \emph {et~al.}(2007)\citenamefont {Reyren},
  \citenamefont {Thiel}, \citenamefont {Caviglia}, \citenamefont {Kourkoutis},
  \citenamefont {Hammerl}, \citenamefont {Richter}, \citenamefont {Schneider},
  \citenamefont {Kopp}, \citenamefont {R\"{u}etschi}, \citenamefont {Jaccard},
  \citenamefont {Gabay}, \citenamefont {Muller}, \citenamefont {Triscone},\
  and\ \citenamefont {Mannhart}}]{Reyren2007}%
  \BibitemOpen
  \bibfield  {author} {\bibinfo {author} {\bibfnamefont {N.}~\bibnamefont
  {Reyren}}, \bibinfo {author} {\bibfnamefont {S.}~\bibnamefont {Thiel}},
  \bibinfo {author} {\bibfnamefont {A.~D.}\ \bibnamefont {Caviglia}}, \bibinfo
  {author} {\bibfnamefont {L.~F.}\ \bibnamefont {Kourkoutis}}, \bibinfo
  {author} {\bibfnamefont {G.}~\bibnamefont {Hammerl}}, \bibinfo {author}
  {\bibfnamefont {C.}~\bibnamefont {Richter}}, \bibinfo {author} {\bibfnamefont
  {C.~W.}\ \bibnamefont {Schneider}}, \bibinfo {author} {\bibfnamefont
  {T.}~\bibnamefont {Kopp}}, \bibinfo {author} {\bibfnamefont {A.-S.}\
  \bibnamefont {R\"{u}etschi}}, \bibinfo {author} {\bibfnamefont
  {D.}~\bibnamefont {Jaccard}}, \bibinfo {author} {\bibfnamefont
  {M.}~\bibnamefont {Gabay}}, \bibinfo {author} {\bibfnamefont {D.~A.}\
  \bibnamefont {Muller}}, \bibinfo {author} {\bibfnamefont {J.-M.}\
  \bibnamefont {Triscone}}, \ and\ \bibinfo {author} {\bibfnamefont
  {J.}~\bibnamefont {Mannhart}},\ }\href {\doibase 10.1126/science.1146006}
  {\bibfield  {journal} {\bibinfo  {journal} {Science}\ }\textbf {\bibinfo
  {volume} {317}},\ \bibinfo {pages} {1196} (\bibinfo {year}
  {2007})}\BibitemShut {NoStop}%
\bibitem [{\citenamefont {Huijben}\ \emph {et~al.}(2009)\citenamefont
  {Huijben}, \citenamefont {Brinkman}, \citenamefont {Koster}, \citenamefont
  {Rijnders}, \citenamefont {Hilgenkamp},\ and\ \citenamefont
  {Blank}}]{Huijben2009}%
  \BibitemOpen
  \bibfield  {author} {\bibinfo {author} {\bibfnamefont {M.}~\bibnamefont
  {Huijben}}, \bibinfo {author} {\bibfnamefont {A.}~\bibnamefont {Brinkman}},
  \bibinfo {author} {\bibfnamefont {G.}~\bibnamefont {Koster}}, \bibinfo
  {author} {\bibfnamefont {G.}~\bibnamefont {Rijnders}}, \bibinfo {author}
  {\bibfnamefont {H.}~\bibnamefont {Hilgenkamp}}, \ and\ \bibinfo {author}
  {\bibfnamefont {D.~H.~A.}\ \bibnamefont {Blank}},\ }\href {\doibase
  10.1002/adma.200801448} {\bibfield  {journal} {\bibinfo  {journal} {Adv.
  Mater.}\ }\textbf {\bibinfo {volume} {21}},\ \bibinfo {pages} {1665}
  (\bibinfo {year} {2009})}\BibitemShut {NoStop}%
\bibitem [{\citenamefont {Thiel}\ \emph {et~al.}(2009)\citenamefont {Thiel},
  \citenamefont {Schneider}, \citenamefont {Kourkoutis}, \citenamefont
  {Muller}, \citenamefont {Reyren}, \citenamefont {Caviglia}, \citenamefont
  {Gariglio}, \citenamefont {Triscone},\ and\ \citenamefont
  {Mannhart}}]{Thiel2009}%
  \BibitemOpen
  \bibfield  {author} {\bibinfo {author} {\bibfnamefont {S.}~\bibnamefont
  {Thiel}}, \bibinfo {author} {\bibfnamefont {C.}~\bibnamefont {Schneider}},
  \bibinfo {author} {\bibfnamefont {L.}~\bibnamefont {Kourkoutis}}, \bibinfo
  {author} {\bibfnamefont {D.}~\bibnamefont {Muller}}, \bibinfo {author}
  {\bibfnamefont {N.}~\bibnamefont {Reyren}}, \bibinfo {author} {\bibfnamefont
  {A.}~\bibnamefont {Caviglia}}, \bibinfo {author} {\bibfnamefont
  {S.}~\bibnamefont {Gariglio}}, \bibinfo {author} {\bibfnamefont {J.-M.}\
  \bibnamefont {Triscone}}, \ and\ \bibinfo {author} {\bibfnamefont
  {J.}~\bibnamefont {Mannhart}},\ }\href {\doibase
  10.1103/PhysRevLett.102.046809} {\bibfield  {journal} {\bibinfo  {journal}
  {Phys. Rev. Lett.}\ }\textbf {\bibinfo {volume} {102}},\ \bibinfo {pages}
  {046809} (\bibinfo {year} {2009})}\BibitemShut {NoStop}%
\bibitem [{\citenamefont {Fix}\ \emph {et~al.}(2011)\citenamefont {Fix},
  \citenamefont {Schoofs}, \citenamefont {Bi}, \citenamefont {Chen},
  \citenamefont {Wang}, \citenamefont {MacManus-Driscoll},\ and\ \citenamefont
  {Blamire}}]{Fix2011}%
  \BibitemOpen
  \bibfield  {author} {\bibinfo {author} {\bibfnamefont {T.}~\bibnamefont
  {Fix}}, \bibinfo {author} {\bibfnamefont {F.}~\bibnamefont {Schoofs}},
  \bibinfo {author} {\bibfnamefont {Z.}~\bibnamefont {Bi}}, \bibinfo {author}
  {\bibfnamefont {A.}~\bibnamefont {Chen}}, \bibinfo {author} {\bibfnamefont
  {H.}~\bibnamefont {Wang}}, \bibinfo {author} {\bibfnamefont {J.~L.}\
  \bibnamefont {MacManus-Driscoll}}, \ and\ \bibinfo {author} {\bibfnamefont
  {M.~G.}\ \bibnamefont {Blamire}},\ }\href {\doibase 10.1063/1.3609785}
  {\bibfield  {journal} {\bibinfo  {journal} {Appl. Phys. Lett.}\ }\textbf
  {\bibinfo {volume} {99}},\ \bibinfo {pages} {022103} (\bibinfo {year}
  {2011})}\BibitemShut {NoStop}%
\bibitem [{\citenamefont {Bell}\ \emph
  {et~al.}(2009{\natexlab{a}})\citenamefont {Bell}, \citenamefont {Harashima},
  \citenamefont {Kozuka}, \citenamefont {Kim}, \citenamefont {Kim},
  \citenamefont {Hikita},\ and\ \citenamefont {Hwang}}]{Bell2009a}%
  \BibitemOpen
  \bibfield  {author} {\bibinfo {author} {\bibfnamefont {C.}~\bibnamefont
  {Bell}}, \bibinfo {author} {\bibfnamefont {S.}~\bibnamefont {Harashima}},
  \bibinfo {author} {\bibfnamefont {Y.}~\bibnamefont {Kozuka}}, \bibinfo
  {author} {\bibfnamefont {M.}~\bibnamefont {Kim}}, \bibinfo {author}
  {\bibfnamefont {B.~G.}\ \bibnamefont {Kim}}, \bibinfo {author} {\bibfnamefont
  {Y.}~\bibnamefont {Hikita}}, \ and\ \bibinfo {author} {\bibfnamefont {H.~Y.}\
  \bibnamefont {Hwang}},\ }\href {\doibase 10.1103/PhysRevLett.103.226802}
  {\bibfield  {journal} {\bibinfo  {journal} {Phys. Rev. Lett.}\ }\textbf
  {\bibinfo {volume} {103}},\ \bibinfo {pages} {226802} (\bibinfo {year}
  {2009}{\natexlab{a}})}\BibitemShut {NoStop}%
\bibitem [{\citenamefont {Kim}\ \emph {et~al.}(2010)\citenamefont {Kim},
  \citenamefont {Seo}, \citenamefont {Chisholm}, \citenamefont {Kremer},
  \citenamefont {Habermeier}, \citenamefont {Keimer},\ and\ \citenamefont
  {Lee}}]{Kim2010}%
  \BibitemOpen
  \bibfield  {author} {\bibinfo {author} {\bibfnamefont {J.~S.}\ \bibnamefont
  {Kim}}, \bibinfo {author} {\bibfnamefont {S.~S.~A.}\ \bibnamefont {Seo}},
  \bibinfo {author} {\bibfnamefont {M.~F.}\ \bibnamefont {Chisholm}}, \bibinfo
  {author} {\bibfnamefont {R.~K.}\ \bibnamefont {Kremer}}, \bibinfo {author}
  {\bibfnamefont {H.-U.}\ \bibnamefont {Habermeier}}, \bibinfo {author}
  {\bibfnamefont {B.}~\bibnamefont {Keimer}}, \ and\ \bibinfo {author}
  {\bibfnamefont {H.~N.}\ \bibnamefont {Lee}},\ }\href {\doibase
  10.1103/PhysRevB.82.201407} {\bibfield  {journal} {\bibinfo  {journal} {Phys.
  Rev. B}\ }\textbf {\bibinfo {volume} {82}},\ \bibinfo {pages} {201407}
  (\bibinfo {year} {2010})}\BibitemShut {NoStop}%
\bibitem [{\citenamefont {Bell}\ \emph
  {et~al.}(2009{\natexlab{b}})\citenamefont {Bell}, \citenamefont {Harashima},
  \citenamefont {Hikita},\ and\ \citenamefont {Hwang}}]{Bell2009}%
  \BibitemOpen
  \bibfield  {author} {\bibinfo {author} {\bibfnamefont {C.}~\bibnamefont
  {Bell}}, \bibinfo {author} {\bibfnamefont {S.}~\bibnamefont {Harashima}},
  \bibinfo {author} {\bibfnamefont {Y.}~\bibnamefont {Hikita}}, \ and\ \bibinfo
  {author} {\bibfnamefont {H.~Y.}\ \bibnamefont {Hwang}},\ }\href {\doibase
  10.1063/1.3149695} {\bibfield  {journal} {\bibinfo  {journal} {Appl. Phys.
  Lett.}\ }\textbf {\bibinfo {volume} {94}},\ \bibinfo {pages} {222111}
  (\bibinfo {year} {2009}{\natexlab{b}})}\BibitemShut {NoStop}%
\bibitem [{\citenamefont {Kalisky}\ \emph {et~al.}(2013)\citenamefont
  {Kalisky}, \citenamefont {Spanton}, \citenamefont {Noad}, \citenamefont
  {Kirtley}, \citenamefont {Nowack}, \citenamefont {Bell}, \citenamefont
  {Sato}, \citenamefont {Hosoda}, \citenamefont {Xie}, \citenamefont {Hikita},
  \citenamefont {Woltmann}, \citenamefont {Pfanzelt}, \citenamefont {Jany},
  \citenamefont {Richter}, \citenamefont {Hwang}, \citenamefont {Mannhart},\
  and\ \citenamefont {Moler}}]{Kalisky2013}%
  \BibitemOpen
  \bibfield  {author} {\bibinfo {author} {\bibfnamefont {B.}~\bibnamefont
  {Kalisky}}, \bibinfo {author} {\bibfnamefont {E.~M.}\ \bibnamefont
  {Spanton}}, \bibinfo {author} {\bibfnamefont {H.}~\bibnamefont {Noad}},
  \bibinfo {author} {\bibfnamefont {J.~R.}\ \bibnamefont {Kirtley}}, \bibinfo
  {author} {\bibfnamefont {K.~C.}\ \bibnamefont {Nowack}}, \bibinfo {author}
  {\bibfnamefont {C.}~\bibnamefont {Bell}}, \bibinfo {author} {\bibfnamefont
  {H.~K.}\ \bibnamefont {Sato}}, \bibinfo {author} {\bibfnamefont
  {M.}~\bibnamefont {Hosoda}}, \bibinfo {author} {\bibfnamefont
  {Y.}~\bibnamefont {Xie}}, \bibinfo {author} {\bibfnamefont {Y.}~\bibnamefont
  {Hikita}}, \bibinfo {author} {\bibfnamefont {C.}~\bibnamefont {Woltmann}},
  \bibinfo {author} {\bibfnamefont {G.}~\bibnamefont {Pfanzelt}}, \bibinfo
  {author} {\bibfnamefont {R.}~\bibnamefont {Jany}}, \bibinfo {author}
  {\bibfnamefont {C.}~\bibnamefont {Richter}}, \bibinfo {author} {\bibfnamefont
  {H.~Y.}\ \bibnamefont {Hwang}}, \bibinfo {author} {\bibfnamefont
  {J.}~\bibnamefont {Mannhart}}, \ and\ \bibinfo {author} {\bibfnamefont
  {K.~A.}\ \bibnamefont {Moler}},\ }\href {\doibase 10.1038/nmat3753}
  {\bibfield  {journal} {\bibinfo  {journal} {Nat. Mater.}\ }\textbf {\bibinfo
  {volume} {12}},\ \bibinfo {pages} {1091} (\bibinfo {year}
  {2013})}\BibitemShut {NoStop}%
\bibitem [{\citenamefont {Honig}\ \emph {et~al.}(2013)\citenamefont {Honig},
  \citenamefont {Sulpizio}, \citenamefont {Drori}, \citenamefont {Joshua},
  \citenamefont {Zeldov},\ and\ \citenamefont {Ilani}}]{Honig2013}%
  \BibitemOpen
  \bibfield  {author} {\bibinfo {author} {\bibfnamefont {M.}~\bibnamefont
  {Honig}}, \bibinfo {author} {\bibfnamefont {J.~a.}\ \bibnamefont {Sulpizio}},
  \bibinfo {author} {\bibfnamefont {J.}~\bibnamefont {Drori}}, \bibinfo
  {author} {\bibfnamefont {A.}~\bibnamefont {Joshua}}, \bibinfo {author}
  {\bibfnamefont {E.}~\bibnamefont {Zeldov}}, \ and\ \bibinfo {author}
  {\bibfnamefont {S.}~\bibnamefont {Ilani}},\ }\href {\doibase
  10.1038/nmat3810} {\bibfield  {journal} {\bibinfo  {journal} {Nat. Mater.}\
  }\textbf {\bibinfo {volume} {12}},\ \bibinfo {pages} {1112} (\bibinfo {year}
  {2013})}\BibitemShut {NoStop}%
\bibitem [{\citenamefont {Xie}\ \emph {et~al.}(2013)\citenamefont {Xie},
  \citenamefont {Bell}, \citenamefont {Hikita}, \citenamefont {Harashima},\
  and\ \citenamefont {Hwang}}]{Xie2013}%
  \BibitemOpen
  \bibfield  {author} {\bibinfo {author} {\bibfnamefont {Y.}~\bibnamefont
  {Xie}}, \bibinfo {author} {\bibfnamefont {C.}~\bibnamefont {Bell}}, \bibinfo
  {author} {\bibfnamefont {Y.}~\bibnamefont {Hikita}}, \bibinfo {author}
  {\bibfnamefont {S.}~\bibnamefont {Harashima}}, \ and\ \bibinfo {author}
  {\bibfnamefont {H.~Y.}\ \bibnamefont {Hwang}},\ }\href {\doibase
  10.1002/adma.201301798} {\bibfield  {journal} {\bibinfo  {journal} {Adv.
  Mater.}\ }\textbf {\bibinfo {volume} {25}},\ \bibinfo {pages} {4735}
  (\bibinfo {year} {2013})}\BibitemShut {NoStop}%
\bibitem [{\citenamefont {Huijben}\ \emph {et~al.}(2013)\citenamefont
  {Huijben}, \citenamefont {Koster}, \citenamefont {Kruize}, \citenamefont
  {Wenderich}, \citenamefont {Verbeeck}, \citenamefont {Bals}, \citenamefont
  {Slooten}, \citenamefont {Shi}, \citenamefont {Molegraaf}, \citenamefont
  {Kleibeuker}, \citenamefont {van Aert}, \citenamefont {Goedkoop},
  \citenamefont {Brinkman}, \citenamefont {Blank}, \citenamefont {Golden},
  \citenamefont {van Tendeloo}, \citenamefont {Hilgenkamp},\ and\ \citenamefont
  {Rijnders}}]{Huijben2013}%
  \BibitemOpen
  \bibfield  {author} {\bibinfo {author} {\bibfnamefont {M.}~\bibnamefont
  {Huijben}}, \bibinfo {author} {\bibfnamefont {G.}~\bibnamefont {Koster}},
  \bibinfo {author} {\bibfnamefont {M.~K.}\ \bibnamefont {Kruize}}, \bibinfo
  {author} {\bibfnamefont {S.}~\bibnamefont {Wenderich}}, \bibinfo {author}
  {\bibfnamefont {J.}~\bibnamefont {Verbeeck}}, \bibinfo {author}
  {\bibfnamefont {S.}~\bibnamefont {Bals}}, \bibinfo {author} {\bibfnamefont
  {E.}~\bibnamefont {Slooten}}, \bibinfo {author} {\bibfnamefont
  {B.}~\bibnamefont {Shi}}, \bibinfo {author} {\bibfnamefont {H.~J.~a.}\
  \bibnamefont {Molegraaf}}, \bibinfo {author} {\bibfnamefont {J.~E.}\
  \bibnamefont {Kleibeuker}}, \bibinfo {author} {\bibfnamefont
  {S.}~\bibnamefont {van Aert}}, \bibinfo {author} {\bibfnamefont {J.~B.}\
  \bibnamefont {Goedkoop}}, \bibinfo {author} {\bibfnamefont {A.}~\bibnamefont
  {Brinkman}}, \bibinfo {author} {\bibfnamefont {D.~H.~a.}\ \bibnamefont
  {Blank}}, \bibinfo {author} {\bibfnamefont {M.~S.}\ \bibnamefont {Golden}},
  \bibinfo {author} {\bibfnamefont {G.}~\bibnamefont {van Tendeloo}}, \bibinfo
  {author} {\bibfnamefont {H.}~\bibnamefont {Hilgenkamp}}, \ and\ \bibinfo
  {author} {\bibfnamefont {G.}~\bibnamefont {Rijnders}},\ }\href {\doibase
  10.1002/adfm.201203355} {\bibfield  {journal} {\bibinfo  {journal} {Adv.
  Funct. Mater.}\ }\textbf {\bibinfo {volume} {23}},\ \bibinfo {pages} {5240}
  (\bibinfo {year} {2013})}\BibitemShut {NoStop}%
\bibitem [{\citenamefont {Caviglia}\ \emph
  {et~al.}(2010{\natexlab{a}})\citenamefont {Caviglia}, \citenamefont
  {Gariglio}, \citenamefont {Cancellieri}, \citenamefont {Sac\'{e}p\'{e}},
  \citenamefont {Fete}, \citenamefont {Reyren}, \citenamefont {Gabay},
  \citenamefont {Morpurgo},\ and\ \citenamefont {Triscone}}]{Caviglia2010105}%
  \BibitemOpen
  \bibfield  {author} {\bibinfo {author} {\bibfnamefont {A.~D.}\ \bibnamefont
  {Caviglia}}, \bibinfo {author} {\bibfnamefont {S.}~\bibnamefont {Gariglio}},
  \bibinfo {author} {\bibfnamefont {C.}~\bibnamefont {Cancellieri}}, \bibinfo
  {author} {\bibfnamefont {B.}~\bibnamefont {Sac\'{e}p\'{e}}}, \bibinfo
  {author} {\bibfnamefont {A.}~\bibnamefont {Fete}}, \bibinfo {author}
  {\bibfnamefont {N.}~\bibnamefont {Reyren}}, \bibinfo {author} {\bibfnamefont
  {M.}~\bibnamefont {Gabay}}, \bibinfo {author} {\bibfnamefont {A.~F.}\
  \bibnamefont {Morpurgo}}, \ and\ \bibinfo {author} {\bibfnamefont {J.-M.}\
  \bibnamefont {Triscone}},\ }\href {\doibase 10.1103/PhysRevLett.105.236802}
  {\bibfield  {journal} {\bibinfo  {journal} {Phys. Rev. Lett.}\ }\textbf
  {\bibinfo {volume} {105}},\ \bibinfo {pages} {236802} (\bibinfo {year}
  {2010}{\natexlab{a}})}\BibitemShut {NoStop}%
\bibitem [{\citenamefont {Cancellieri}\ \emph {et~al.}(2010)\citenamefont
  {Cancellieri}, \citenamefont {Reyren}, \citenamefont {Gariglio},
  \citenamefont {Caviglia}, \citenamefont {Fete},\ and\ \citenamefont
  {Triscone}}]{Cancellieri2010}%
  \BibitemOpen
  \bibfield  {author} {\bibinfo {author} {\bibfnamefont {C.}~\bibnamefont
  {Cancellieri}}, \bibinfo {author} {\bibfnamefont {N.}~\bibnamefont {Reyren}},
  \bibinfo {author} {\bibfnamefont {S.}~\bibnamefont {Gariglio}}, \bibinfo
  {author} {\bibfnamefont {A.~D.}\ \bibnamefont {Caviglia}}, \bibinfo {author}
  {\bibfnamefont {A.}~\bibnamefont {Fete}}, \ and\ \bibinfo {author}
  {\bibfnamefont {J.-M.}\ \bibnamefont {Triscone}},\ }\href {\doibase
  10.1209/0295-5075/91/17004} {\bibfield  {journal} {\bibinfo  {journal} {EPL
  (Europhysics Lett.}\ }\textbf {\bibinfo {volume} {91}},\ \bibinfo {pages}
  {17004} (\bibinfo {year} {2010})}\BibitemShut {NoStop}%
\bibitem [{Note1()}]{Note1}%
  \BibitemOpen
  \bibinfo {note} {We use STO substrates from CrysTec GmbH that are already HF
  treated and high-temperature annealed to provide \protect \mhchem@ce@xiii
  {\protect \noexpand \protect T\protect \noexpand \protect i\protect \noexpand
  \protect O\protect \noexpand \protect 2} termination. Before growth, the
  sample holder with the substrate is annealed in vacuum for 30 min at a
  temperature of \SI {290}{\celsius }.}\BibitemShut {Stop}%
\bibitem [{\citenamefont {Stornaiuolo}\ \emph {et~al.}(2012)\citenamefont
  {Stornaiuolo}, \citenamefont {Gariglio}, \citenamefont {Couto}, \citenamefont
  {F\^ete}, \citenamefont {Caviglia}, \citenamefont {Seyfarth}, \citenamefont
  {Jaccard}, \citenamefont {Morpurgo},\ and\ \citenamefont
  {Triscone}}]{Stornaiuolo2012}%
  \BibitemOpen
  \bibfield  {author} {\bibinfo {author} {\bibfnamefont {D.}~\bibnamefont
  {Stornaiuolo}}, \bibinfo {author} {\bibfnamefont {S.}~\bibnamefont
  {Gariglio}}, \bibinfo {author} {\bibfnamefont {N.~J.~G.}\ \bibnamefont
  {Couto}}, \bibinfo {author} {\bibfnamefont {A.}~\bibnamefont {F\^ete}},
  \bibinfo {author} {\bibfnamefont {a.~D.}\ \bibnamefont {Caviglia}}, \bibinfo
  {author} {\bibfnamefont {G.}~\bibnamefont {Seyfarth}}, \bibinfo {author}
  {\bibfnamefont {D.}~\bibnamefont {Jaccard}}, \bibinfo {author} {\bibfnamefont
  {a.~F.}\ \bibnamefont {Morpurgo}}, \ and\ \bibinfo {author} {\bibfnamefont
  {J.-M.}\ \bibnamefont {Triscone}},\ }\href {\doibase 10.1063/1.4768936}
  {\bibfield  {journal} {\bibinfo  {journal} {Appl. Phys. Lett.}\ }\textbf
  {\bibinfo {volume} {101}},\ \bibinfo {pages} {222601} (\bibinfo {year}
  {2012})}\BibitemShut {NoStop}%
\bibitem [{Note2()}]{Note2}%
  \BibitemOpen
  \bibinfo {note} {The estimation of the carrier density is obtained from the
  Hall constant at \SI {296}{\kelvin } where the Hall resistance is
  linear.}\BibitemShut {Stop}%
\bibitem [{Note3()}]{Note3}%
  \BibitemOpen
  \bibinfo {note} {We note, though, that depending on the photo-doping level
  \cite {Tebano2012,Guduru2013}, some temperature dependence of the Hall effect
  above \SI {100}{\kelvin } can sometimes be observed.}\BibitemShut {Stop}%
\bibitem [{\citenamefont {{Ben Shalom}}\ \emph {et~al.}(2010)\citenamefont
  {{Ben Shalom}}, \citenamefont {Ron}, \citenamefont {Palevski},\ and\
  \citenamefont {Dagan}}]{BenShalom2010105}%
  \BibitemOpen
  \bibfield  {author} {\bibinfo {author} {\bibfnamefont {M.}~\bibnamefont {{Ben
  Shalom}}}, \bibinfo {author} {\bibfnamefont {A.}~\bibnamefont {Ron}},
  \bibinfo {author} {\bibfnamefont {A.}~\bibnamefont {Palevski}}, \ and\
  \bibinfo {author} {\bibfnamefont {Y.}~\bibnamefont {Dagan}},\ }\href
  {\doibase 10.1103/PhysRevLett.105.206401} {\bibfield  {journal} {\bibinfo
  {journal} {Phys. Rev. Lett.}\ }\textbf {\bibinfo {volume} {105}},\ \bibinfo
  {pages} {206401} (\bibinfo {year} {2010})}\BibitemShut {NoStop}%
\bibitem [{\citenamefont {Lerer}\ \emph {et~al.}(2011)\citenamefont {Lerer},
  \citenamefont {{Ben Shalom}}, \citenamefont {Deutscher},\ and\ \citenamefont
  {Dagan}}]{Lerer2011}%
  \BibitemOpen
  \bibfield  {author} {\bibinfo {author} {\bibfnamefont {S.}~\bibnamefont
  {Lerer}}, \bibinfo {author} {\bibfnamefont {M.}~\bibnamefont {{Ben Shalom}}},
  \bibinfo {author} {\bibfnamefont {G.}~\bibnamefont {Deutscher}}, \ and\
  \bibinfo {author} {\bibfnamefont {Y.}~\bibnamefont {Dagan}},\ }\href
  {\doibase 10.1103/PhysRevB.84.075423} {\bibfield  {journal} {\bibinfo
  {journal} {Phys. Rev. B}\ }\textbf {\bibinfo {volume} {84}},\ \bibinfo
  {pages} {075423} (\bibinfo {year} {2011})}\BibitemShut {NoStop}%
\bibitem [{Note4()}]{Note4}%
  \BibitemOpen
  \bibinfo {note} {The elastic scattering time for each band is calculated from
  the respective mobility obtained from the two band fit in magnetic field
  while the inelastic scattering time is considered to be independent off the
  carrier type.}\BibitemShut {Stop}%
\bibitem [{\citenamefont {Takahashi}\ \emph {et~al.}(2006)\citenamefont
  {Takahashi}, \citenamefont {Gabay}, \citenamefont {Jaccard}, \citenamefont
  {Shibuya}, \citenamefont {Ohnishi}, \citenamefont {Lippmaa},\ and\
  \citenamefont {Triscone}}]{Takahashi2006}%
  \BibitemOpen
  \bibfield  {author} {\bibinfo {author} {\bibfnamefont {K.~S.}\ \bibnamefont
  {Takahashi}}, \bibinfo {author} {\bibfnamefont {M.}~\bibnamefont {Gabay}},
  \bibinfo {author} {\bibfnamefont {D.}~\bibnamefont {Jaccard}}, \bibinfo
  {author} {\bibfnamefont {K.}~\bibnamefont {Shibuya}}, \bibinfo {author}
  {\bibfnamefont {T.}~\bibnamefont {Ohnishi}}, \bibinfo {author} {\bibfnamefont
  {M.}~\bibnamefont {Lippmaa}}, \ and\ \bibinfo {author} {\bibfnamefont
  {J.-M.}\ \bibnamefont {Triscone}},\ }\href {\doibase 10.1038/nature04731}
  {\bibfield  {journal} {\bibinfo  {journal} {Nature}\ }\textbf {\bibinfo
  {volume} {441}},\ \bibinfo {pages} {195} (\bibinfo {year}
  {2006})}\BibitemShut {NoStop}%
\bibitem [{\citenamefont {Pentcheva}\ and\ \citenamefont
  {Pickett}(2006)}]{Pentcheva2006}%
  \BibitemOpen
  \bibfield  {author} {\bibinfo {author} {\bibfnamefont {R.}~\bibnamefont
  {Pentcheva}}\ and\ \bibinfo {author} {\bibfnamefont {W.}~\bibnamefont
  {Pickett}},\ }\href {\doibase 10.1103/PhysRevB.74.035112} {\bibfield
  {journal} {\bibinfo  {journal} {Phys. Rev. B}\ }\textbf {\bibinfo {volume}
  {74}},\ \bibinfo {pages} {035112} (\bibinfo {year} {2006})}\BibitemShut
  {NoStop}%
\bibitem [{\citenamefont {Popovi\'{c}}\ \emph {et~al.}(2008)\citenamefont
  {Popovi\'{c}}, \citenamefont {Satpathy},\ and\ \citenamefont
  {Martin}}]{Popovic2008}%
  \BibitemOpen
  \bibfield  {author} {\bibinfo {author} {\bibfnamefont {Z.}~\bibnamefont
  {Popovi\'{c}}}, \bibinfo {author} {\bibfnamefont {S.}~\bibnamefont
  {Satpathy}}, \ and\ \bibinfo {author} {\bibfnamefont {R.}~\bibnamefont
  {Martin}},\ }\href {\doibase 10.1103/PhysRevLett.101.256801} {\bibfield
  {journal} {\bibinfo  {journal} {Phys. Rev. Lett.}\ }\textbf {\bibinfo
  {volume} {101}},\ \bibinfo {pages} {256801} (\bibinfo {year}
  {2008})}\BibitemShut {NoStop}%
\bibitem [{\citenamefont {Son}\ \emph {et~al.}(2009)\citenamefont {Son},
  \citenamefont {Cho}, \citenamefont {Lee}, \citenamefont {Lee},\ and\
  \citenamefont {Han}}]{Son2009}%
  \BibitemOpen
  \bibfield  {author} {\bibinfo {author} {\bibfnamefont {W.-j.}\ \bibnamefont
  {Son}}, \bibinfo {author} {\bibfnamefont {E.}~\bibnamefont {Cho}}, \bibinfo
  {author} {\bibfnamefont {B.}~\bibnamefont {Lee}}, \bibinfo {author}
  {\bibfnamefont {J.}~\bibnamefont {Lee}}, \ and\ \bibinfo {author}
  {\bibfnamefont {S.}~\bibnamefont {Han}},\ }\href {\doibase
  10.1103/PhysRevB.79.245411} {\bibfield  {journal} {\bibinfo  {journal} {Phys.
  Rev. B}\ }\textbf {\bibinfo {volume} {79}},\ \bibinfo {pages} {245411}
  (\bibinfo {year} {2009})}\BibitemShut {NoStop}%
\bibitem [{\citenamefont {Delugas}\ \emph {et~al.}(2011)\citenamefont
  {Delugas}, \citenamefont {Filippetti}, \citenamefont {Fiorentini},
  \citenamefont {Bilc}, \citenamefont {Fontaine},\ and\ \citenamefont
  {Ghosez}}]{Delugas2011}%
  \BibitemOpen
  \bibfield  {author} {\bibinfo {author} {\bibfnamefont {P.}~\bibnamefont
  {Delugas}}, \bibinfo {author} {\bibfnamefont {A.}~\bibnamefont {Filippetti}},
  \bibinfo {author} {\bibfnamefont {V.}~\bibnamefont {Fiorentini}}, \bibinfo
  {author} {\bibfnamefont {D.~I.}\ \bibnamefont {Bilc}}, \bibinfo {author}
  {\bibfnamefont {D.}~\bibnamefont {Fontaine}}, \ and\ \bibinfo {author}
  {\bibfnamefont {P.}~\bibnamefont {Ghosez}},\ }\href {\doibase
  10.1103/PhysRevLett.106.166807} {\bibfield  {journal} {\bibinfo  {journal}
  {Phys. Rev. Lett.}\ }\textbf {\bibinfo {volume} {106}},\ \bibinfo {pages}
  {166807} (\bibinfo {year} {2011})}\BibitemShut {NoStop}%
\bibitem [{\citenamefont {Salluzzo}\ \emph {et~al.}(2009)\citenamefont
  {Salluzzo}, \citenamefont {Cezar}, \citenamefont {Brookes}, \citenamefont
  {Bisogni}, \citenamefont {{De Luca}}, \citenamefont {Richter}, \citenamefont
  {Thiel}, \citenamefont {Mannhart}, \citenamefont {Huijben}, \citenamefont
  {Brinkman}, \citenamefont {Rijnders},\ and\ \citenamefont
  {Ghiringhelli}}]{Salluzzo2009}%
  \BibitemOpen
  \bibfield  {author} {\bibinfo {author} {\bibfnamefont {M.}~\bibnamefont
  {Salluzzo}}, \bibinfo {author} {\bibfnamefont {J.}~\bibnamefont {Cezar}},
  \bibinfo {author} {\bibfnamefont {N.}~\bibnamefont {Brookes}}, \bibinfo
  {author} {\bibfnamefont {V.}~\bibnamefont {Bisogni}}, \bibinfo {author}
  {\bibfnamefont {G.}~\bibnamefont {{De Luca}}}, \bibinfo {author}
  {\bibfnamefont {C.}~\bibnamefont {Richter}}, \bibinfo {author} {\bibfnamefont
  {S.}~\bibnamefont {Thiel}}, \bibinfo {author} {\bibfnamefont
  {J.}~\bibnamefont {Mannhart}}, \bibinfo {author} {\bibfnamefont
  {M.}~\bibnamefont {Huijben}}, \bibinfo {author} {\bibfnamefont
  {A.}~\bibnamefont {Brinkman}}, \bibinfo {author} {\bibfnamefont
  {G.}~\bibnamefont {Rijnders}}, \ and\ \bibinfo {author} {\bibfnamefont
  {G.}~\bibnamefont {Ghiringhelli}},\ }\href {\doibase
  10.1103/PhysRevLett.102.166804} {\bibfield  {journal} {\bibinfo  {journal}
  {Phys. Rev. Lett.}\ }\textbf {\bibinfo {volume} {102}},\ \bibinfo {pages}
  {166804} (\bibinfo {year} {2009})}\BibitemShut {NoStop}%
\bibitem [{\citenamefont {Cancellieri}\ \emph {et~al.}(2014)\citenamefont
  {Cancellieri}, \citenamefont {Reinle-Schmitt}, \citenamefont {Kobayashi},
  \citenamefont {Strocov}, \citenamefont {Willmott}, \citenamefont {Fontaine},
  \citenamefont {Ghosez}, \citenamefont {Filippetti}, \citenamefont {Delugas},\
  and\ \citenamefont {Fiorentini}}]{Cancellieri2014}%
  \BibitemOpen
  \bibfield  {author} {\bibinfo {author} {\bibfnamefont {C.}~\bibnamefont
  {Cancellieri}}, \bibinfo {author} {\bibfnamefont {M.~L.}\ \bibnamefont
  {Reinle-Schmitt}}, \bibinfo {author} {\bibfnamefont {M.}~\bibnamefont
  {Kobayashi}}, \bibinfo {author} {\bibfnamefont {V.~N.}\ \bibnamefont
  {Strocov}}, \bibinfo {author} {\bibfnamefont {P.~R.}\ \bibnamefont
  {Willmott}}, \bibinfo {author} {\bibfnamefont {D.}~\bibnamefont {Fontaine}},
  \bibinfo {author} {\bibfnamefont {P.}~\bibnamefont {Ghosez}}, \bibinfo
  {author} {\bibfnamefont {a.}~\bibnamefont {Filippetti}}, \bibinfo {author}
  {\bibfnamefont {P.}~\bibnamefont {Delugas}}, \ and\ \bibinfo {author}
  {\bibfnamefont {V.}~\bibnamefont {Fiorentini}},\ }\href {\doibase
  10.1103/PhysRevB.89.121412} {\bibfield  {journal} {\bibinfo  {journal} {Phys.
  Rev. B}\ }\textbf {\bibinfo {volume} {89}},\ \bibinfo {pages} {121412}
  (\bibinfo {year} {2014})}\BibitemShut {NoStop}%
\bibitem [{\citenamefont {F\^ete}\ \emph {et~al.}(2014)\citenamefont {F\^ete},
  \citenamefont {Gariglio}, \citenamefont {Berthod}, \citenamefont {Li},
  \citenamefont {Stornaiuolo}, \citenamefont {Gabay},\ and\ \citenamefont
  {Triscone}}]{Fete2014}%
  \BibitemOpen
  \bibfield  {author} {\bibinfo {author} {\bibfnamefont {A.}~\bibnamefont
  {F\^ete}}, \bibinfo {author} {\bibfnamefont {S.}~\bibnamefont {Gariglio}},
  \bibinfo {author} {\bibfnamefont {C.}~\bibnamefont {Berthod}}, \bibinfo
  {author} {\bibfnamefont {D.}~\bibnamefont {Li}}, \bibinfo {author}
  {\bibfnamefont {D.}~\bibnamefont {Stornaiuolo}}, \bibinfo {author}
  {\bibfnamefont {M.}~\bibnamefont {Gabay}}, \ and\ \bibinfo {author}
  {\bibfnamefont {J.-M.}\ \bibnamefont {Triscone}},\ }\href {\doibase
  10.1088/1367-2630/16/11/112002} {\bibfield  {journal} {\bibinfo  {journal}
  {New J. Phys.}\ }\textbf {\bibinfo {volume} {16}},\ \bibinfo {pages} {112002}
  (\bibinfo {year} {2014})}\BibitemShut {NoStop}%
\bibitem [{Note5()}]{Note5}%
  \BibitemOpen
  \bibinfo {note} {In this carrier density range, the Hall effect is linear in
  magnetic field for both type of interfaces.}\BibitemShut {Stop}%
\bibitem [{\citenamefont {Caviglia}\ \emph
  {et~al.}(2010{\natexlab{b}})\citenamefont {Caviglia}, \citenamefont {Gabay},
  \citenamefont {Gariglio}, \citenamefont {Reyren}, \citenamefont
  {Cancellieri},\ and\ \citenamefont {Triscone}}]{Caviglia2010}%
  \BibitemOpen
  \bibfield  {author} {\bibinfo {author} {\bibfnamefont {A.~D.}\ \bibnamefont
  {Caviglia}}, \bibinfo {author} {\bibfnamefont {M.}~\bibnamefont {Gabay}},
  \bibinfo {author} {\bibfnamefont {S.}~\bibnamefont {Gariglio}}, \bibinfo
  {author} {\bibfnamefont {N.}~\bibnamefont {Reyren}}, \bibinfo {author}
  {\bibfnamefont {C.}~\bibnamefont {Cancellieri}}, \ and\ \bibinfo {author}
  {\bibfnamefont {J.-M.}\ \bibnamefont {Triscone}},\ }\href {\doibase
  10.1103/PhysRevLett.104.126803} {\bibfield  {journal} {\bibinfo  {journal}
  {Phys. Rev. Lett.}\ }\textbf {\bibinfo {volume} {104}},\ \bibinfo {pages}
  {126803} (\bibinfo {year} {2010}{\natexlab{b}})}\BibitemShut {NoStop}%
\bibitem [{\citenamefont {Ohnishi}\ \emph {et~al.}(2004)\citenamefont
  {Ohnishi}, \citenamefont {Shibuya}, \citenamefont {Lippmaa}, \citenamefont
  {Kobayashi}, \citenamefont {Kumigashira}, \citenamefont {Oshima},\ and\
  \citenamefont {Koinuma}}]{Ohnishi2004}%
  \BibitemOpen
  \bibfield  {author} {\bibinfo {author} {\bibfnamefont {T.}~\bibnamefont
  {Ohnishi}}, \bibinfo {author} {\bibfnamefont {K.}~\bibnamefont {Shibuya}},
  \bibinfo {author} {\bibfnamefont {M.}~\bibnamefont {Lippmaa}}, \bibinfo
  {author} {\bibfnamefont {D.}~\bibnamefont {Kobayashi}}, \bibinfo {author}
  {\bibfnamefont {H.}~\bibnamefont {Kumigashira}}, \bibinfo {author}
  {\bibfnamefont {M.}~\bibnamefont {Oshima}}, \ and\ \bibinfo {author}
  {\bibfnamefont {H.}~\bibnamefont {Koinuma}},\ }\href {\doibase
  10.1063/1.1771461} {\bibfield  {journal} {\bibinfo  {journal} {Appl. Phys.
  Lett.}\ }\textbf {\bibinfo {volume} {85}},\ \bibinfo {pages} {272} (\bibinfo
  {year} {2004})}\BibitemShut {NoStop}%
\end{thebibliography}

\end{document}